\documentclass[12pt]{iopart}
\usepackage{epsfig}
\usepackage{graphicx}
\begin{document}

\title{A single hollow beam optical trap for cold atoms}

\author{S Kulin\dag\ S Aubin\ddag S Christe\ddag B Peker\ddag S L Rolston\dag
and L A Orozco\ddag}

\address{\dag\ National Institute of Standards and Technology,
Gaithersburg, MD 20899-8424, USA}

\address{\ddag\ Department of Physics and Astronomy,
State University of New York at Stony Brook, Stony Brook, NY
11794-3800, USA\\ E-mail: Luis.Orozco@SUNYSB.EDU}

\begin{abstract}
We present an optical trap for atoms that we have developed for
precision spectroscopy measurements. Cold atoms are captured in a
dark region of space inside a blue-detuned hollow laser beam
formed by an axicon. We analyze the light potential in a ray
optics picture and experimentally demonstrate trapping of
laser-cooled metastable xenon atoms.

\end{abstract}
\maketitle
\section{Introduction}

Precision spectroscopy and experiments to test discrete
fundamental symmetries in atomic systems, such as Parity or Time,
benefit from long coherence times and perturbation-free
environments. Blue-detuned optical dipole traps, in which atoms
are confined to minima of the light field, may meet both of these
requirements: Trapping insures long interrogation times of the
atoms, while the low light level, or sometimes the absence of
light in the trapping region reduces perturbations from the trap
itself.

There are several realizations of such blue-detuned optical traps
in the literature, and the geometrical and optical arrangements
vary considerably. The first traps of this kind used multiple
laser beams to provide a closed trapping volume for the atoms.
Sodium atoms were captured in between sheets of laser light
\cite{davidson95}, cesium atoms were stored in a trap that uses
gravity and a combination of an evanescent wave and a hollow beam
\cite{grimm}, and rubidium atoms were trapped inside a doughnut
beam closed by additional laser beams along the direction of
propagation \cite{kuga97}. More recently, three far off resonant
traps for cold atoms that use a single laser beam were
demonstrated experimentally. In one case the dark trapping region
was achieved by use of a phase plate that introduces a phase
difference of $\pi$ between the central and outer part of the
laser beam \cite{ozeri99}. In the other two experiments a rotating
focused Gaussian beam produced the confining potential that
enclosed a light-free volume \cite{lev,nick}.

We present in this article a new single beam optical dipole trap
\cite{naples}, which we have developed for possible use in a
measurement of parity non-conservation (PNC) in francium
\cite{simsarian99}. The cold atoms are trapped inside a hollow beam which
is obtained by
placing an axicon or conical lens into the path of the laser light
that forms the ``axicon trap''. Because in PNC experiments the signals are
generally
rather small, we are interested in a large trapping volume. In the
case of the axicon trap the dark volume can be as large as
$80$\,mm$^{3}$ and we have trapped up to one million atoms in it.

This  paper is organized as follows.
We first discuss the formation of the region of darkness using ray
optics (section II). Predictions are compared to measurements of the
intensity at different positions along the optical axis.
In section III we present the experimental implementation of the axicon
trap using a sample of precooled metastable xenon atoms.
Finally, we comment on conditions and usage of the axicon trap for
PNC experiments in francium.

\section{Ray optics of the axicon trap}

In a far-off-resonance blue-detuned optical dipole trap the atoms
experience a confining potential and may spend most of their time
in the dark, provided the trap is designed appropriately. The
simplest implementation of such a trap uses a single hollow laser
beam. The atoms are captured in a region of darkness that is
completely surrounded by light. We realize this confining
``black-box'' by means of an axicon. The use of axicons in atom
manipulation is well established \cite{grimm,golub90,song99}, as
they are a way to produce hollow beams of light.

\begin{figure}[h]
\leavevmode
\centering
\epsfxsize=14cm
\epsffile{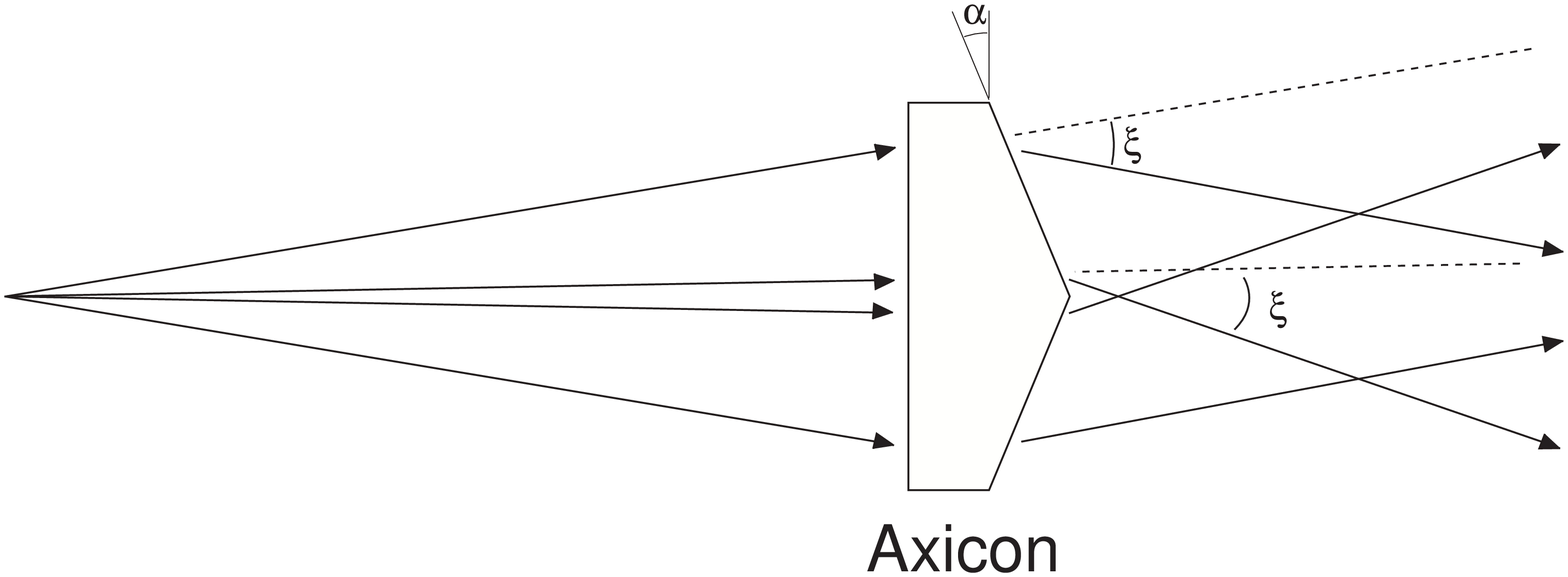}
\caption{Ray optics diagram of the effect of an axicon on a
diverging beam.} \label{fig1}
\end{figure}
Axicons, originally proposed by McLeod \cite{mcleod54}, have also
been studied for imaging systems, and to produce Bessel beams
\cite{soroko89,parigger97,belanger76,belanger78}. In contrast to
an ordinary lens which has a spherical surface, the axicon has a
conical surface. For an apex angle of $\pi-2\alpha$ and an index
of refraction $n$, all incident rays are deviated by the same
angle $\xi=\alpha(n-1)$. Fig.~\ref{fig1} illustrates the path of a
diverging light beam through an axicon lens. The incident beam is
split into two beams of opposite inclinations.

\begin{figure}[h]
\leavevmode
\centering
\epsfxsize=14cm
\epsffile{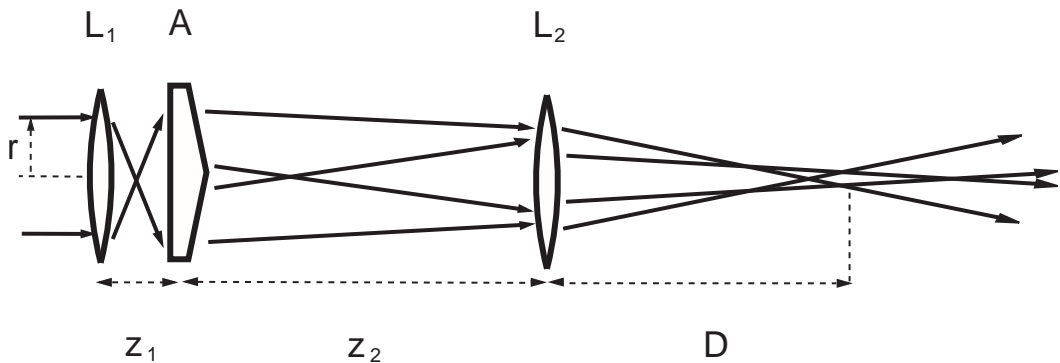}
\caption{Optical setup and ray diagram used to obtain the region
of darkness with the axicon. The drawing has different horizontal
and vertical scales. The horizontal ratio of spacings is that for
the experimental parameters used. \label{fig2}}
\end{figure}

\begin{figure}
\leavevmode
\centering
\epsfxsize=14cm
\epsffile{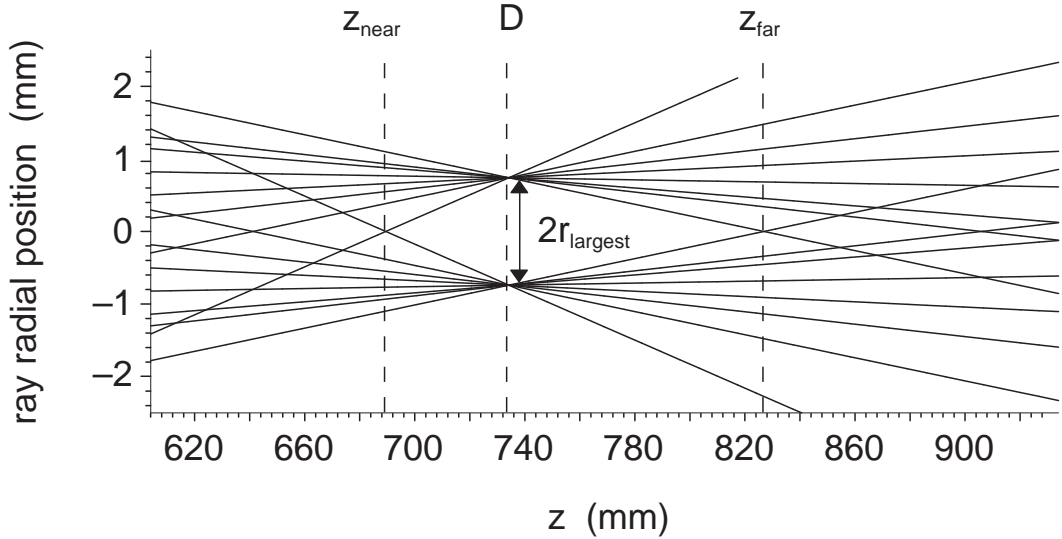}
\caption{Ray tracing at the darkness region that forms behind the
lens $L_{2}$ for $f_1=50.8$~mm, $z_1=163$~mm, $f_2=405$~mm,
$z_2=799$~mm, $\xi= 8.2$ mrad. The scales of the two axis are
different.} \label{fig3}
\end{figure}

We use the optical setup shown in Fig.~\ref{fig2} in order to
produce a region of darkness enclosed by light. It employs two
spherical lenses $L_{1}$ and $L_{2}$ of focal lengths $f_{1}$ and
$f_{2}$ respectively, in addition to the axicon $A$. The two
lenses are placed at a distance $z_{1}$ before and $z_{2}$ behind
the axicon. We denote $r$ the radius of the light beam that enters
the setup at lens $L_{1}$.
When a parallel light beam enters the setup a region of darkness
forms beyond the second lens. A simple ray optics diagram
(Fig.~\ref{fig3}) illustrates the formation of this region in
which no rays cross. The conical lens focuses incident diverging
light into a circle, rather than a spot as would be the case for a
spherical lens. The distance $D$ from the second lens at which the
circle of foci forms, in the paraxial and thin lens approximation
is given by:
\begin{equation}
D=f_2\frac{z_1+z_2-f_1}{z_1+z_2-(f_1+f_2)}.
\label{distanceD}
\end{equation}
The position of the focal plane of the setup, at $D$, is
independent of the presence of the axicon.

Ray tracing allows us to determine which rays cross in order to
form the cusps that appear at the closer ($z=z_{\rm{near}}$) and
distant ($z=z_{\rm {far}}$) boundaries of the dark region. Rays
originating from the outer edge of the incident parallel beam
cross at $z_{\rm{near}}$, whereas incident rays very close to the
optical axis (when $r\rightarrow 0$) form the cusp at
$z_{\rm{far}}$. With this in mind, we find the positions of the
cusps (measured from the lens $L_{2}$, provided that $z_1 > f_1,
z_2 > f_2$):
\begin{equation}
z_{\rm near}=\frac{r\Bigr(1-\frac{z_1}{f_1}\Bigl)-z_{2} \Bigr(-\xi
+ \frac{r}{f_1} \Bigl)} {\frac{r}{f_2} \Bigr(1-\frac{z_1}{f_1}
\Bigl) + \Bigr(1 - \frac{z_2}{f_2} \Bigl) \Bigr( -\xi +
\frac{r}{f_1} \Bigl)},
\label{close}
\end{equation}
and
\begin{equation}
\frac{1}{z_{\rm far}}=\frac{1}{f_2}-\frac{1}{z_2}.
\label{far}
\end{equation}

In order to calculate the total volume of the dark region we need to
determine the radius of the circle of foci. Again, using only
geometric optics
we find that the largest transverse size of the dark region  (at $z=D$ )
is given by:
\begin{equation}
r_{\rm largest}=\frac{\xi f_2(z_1-f_1)}{z_1+z_2-(f_1+f_2)}.
\label{radial}
\end{equation}
This equation shows that the size of the circle of foci
depends directly on the deflection angle $\xi$ and thus the angle
$\alpha$ of the axicon.

The ray optics calculations can predict the shape and the location
of the dark region in space, but cannot give any information about
the size of the beam at the focal plane, or more precisely, the
width of the ring of foci $w_{ring}$. We can, however, determine
the divergence angle of the beam at the circle of foci. It equals
half the divergence angle of the beam that would be focused at
$z=D$ in absence of the axicon lens. In the large distance limit
of Gaussian optics the waist at the focus is inversely
proportional to the divergence angle. Therefore
$w_{ring}=2w_{spot}$, where $w_{spot}$ designates the waist of the
beam formed without the axicon. This result agrees with
calculations of Refs. \cite{parigger97,belanger76,belanger78}.

\begin{figure}[h]
\leavevmode
\centering
\epsfxsize=14cm
\epsffile{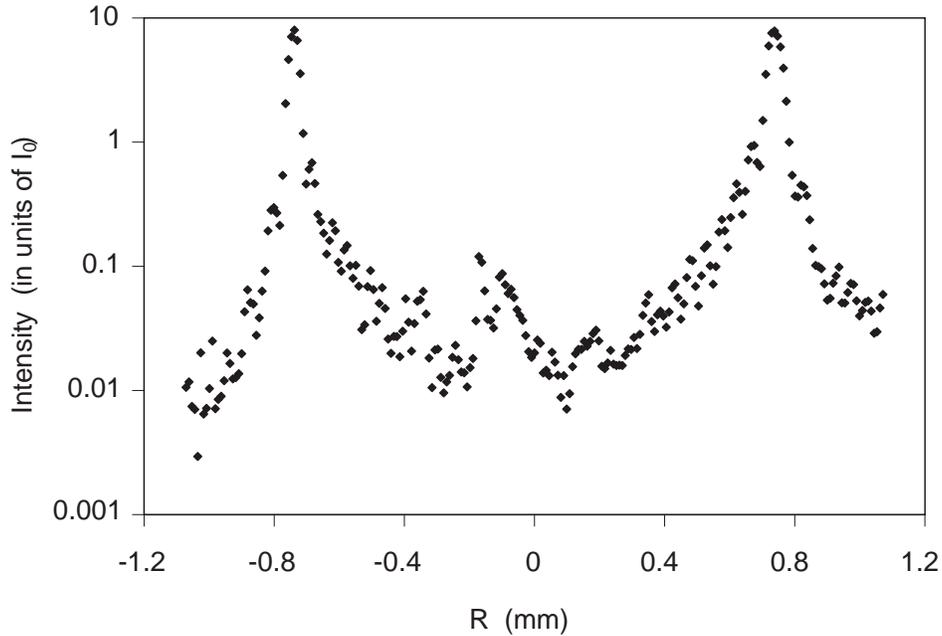}
\caption{Normalized radial intensity distribution of the trapping
light at the location of the circle of foci $z=D$. The measured FWHM of the focal ring is $51\,\mu$m.}
\label{transverse}
\end{figure}

\begin{figure}
\leavevmode \centering
\includegraphics[width=14cm]{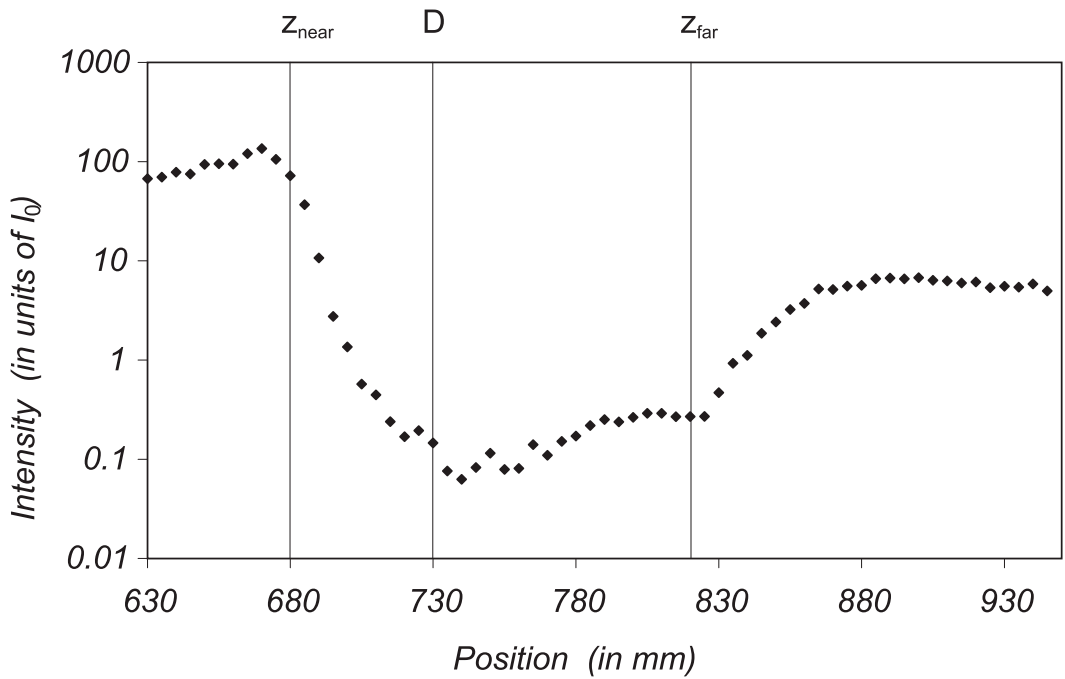} \caption{On-axis intensity of the
light. The distance $z$ is measured from the lens $L_{\rm 2}$. The
vertical lines indicate the location of the calculated values for
$z_{\rm near}$, $z_{\rm far}$ and $D$.} \label{longitudinal}
\end{figure}

We measured the intensity profile of the laser beam shaped by the
optical setup of Fig.~\ref{fig2} by using a charged coupled device
(CCD) camera with 16 bit resolution. The parameters used in the
experiment are: $f_1=50.8$~mm, $f_2=405$~mm, $z_1=163$~mm,
$z_2=799$~mm. In an independent measurement we determined
$\alpha=18.2 \pm 0.4$ mrad, which leads to $\xi= 8.2\pm 0.2$ mrad.
The trap region is an elongated diamond of revolution with a
diameter of roughly 1.5 mm and 150 mm of length. The measured
intensity distributions in the radial direction, at the location
of the circle of foci ($z=D$), and along the optical axis are
shown in figures~\ref{transverse} and ~\ref{longitudinal},
respectively. The recorded intensities are normalized to $I_0$,
the intensity at lens $L_1$. The location of the circle of foci
$D_{\rm measured}=730 \pm 3$\, mm agrees with the theoretical
calculation from Eq.~(\ref{distanceD}) of $D_{\rm theory}=730 $mm.
The measured distance between the two maxima, $2r_{\rm largest,
measured}=1.48\pm 0.02$\,mm, is in accord with the prediction of
the diameter of the circle of foci $2r_{\rm largest,
theory}=1.479$\,mm (eq.~(\ref{radial})).

In Fig.~\ref{longitudinal} the calculated locations of the cusp
positions $z_{\rm near}=688$ mm and $z_{\rm far}=822$ mm are
indicated. Because in the vicinity of these points the intensity
does not increase sharply, it is difficult to clearly identify
$z_{\rm near}$ and $z_{\rm far}$ in the measured intensity
distribution. Diffraction smoothes the onset of the intensity
maxima and imperfections of the axicon lens can further
deteriorate the dark region. Because the axicon is not a perfect
cone, and in particular when it has a slightly flattened apex,
light may pass directly on axis. By approximating the trapping
volume with two cones that share a base we infer a volume of $\sim
80$\,mm$^{3}$ in which atoms may be captured.

\begin{figure}
\leavevmode \centering \epsfxsize=14cm \epsffile{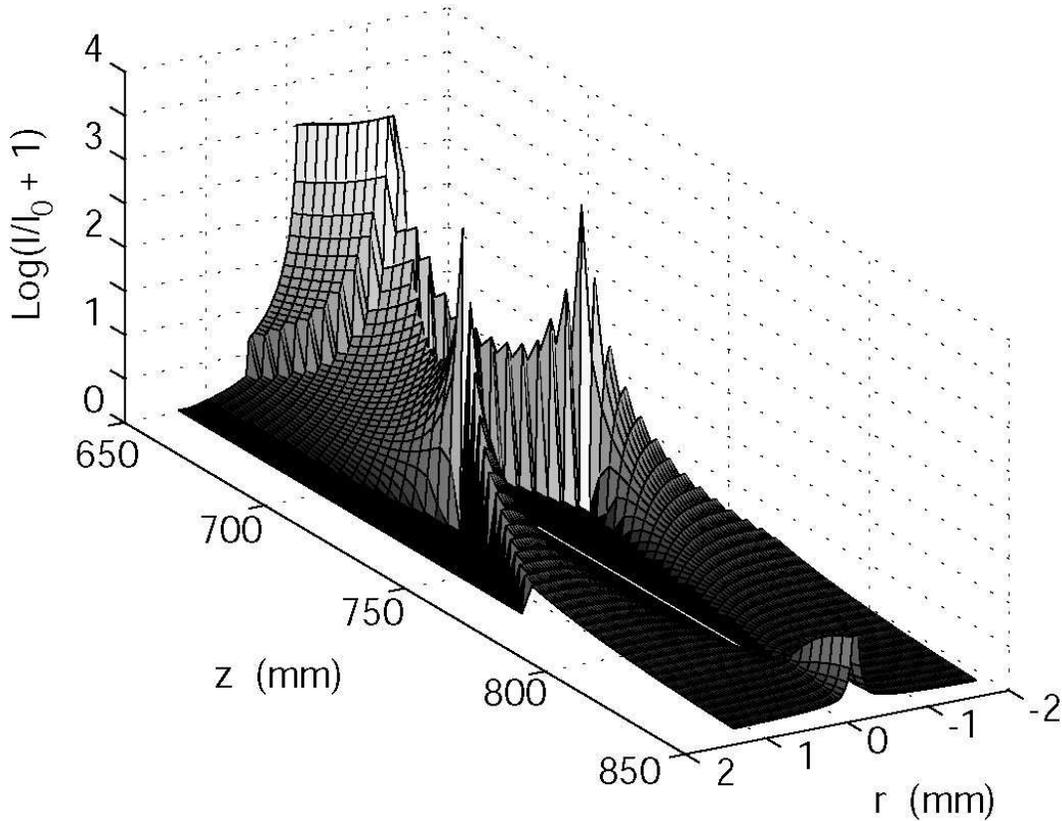}
 \caption{Normalized
intensity of the light around the trapping region from geometrical
optics calculations for $f_1=50.8$\,mm, $f_2=405$\,mm,
$z_1=163$\,mm, and $z_2=799$\,mm. Since the calculation produces
ample regions of zeroes, we added one to facilitate the depiction
in a logarithmic scale. } \label{3d}
\end{figure}

The optical potential $U({\bf r})$ seen by the trapped atoms is
directly proportional to the light intensity $I({\bf r})$
\cite{metcalf99}, for large detuning from the optical transition.
Figure \ref{3d} shows a three dimensional rendition of the axial
and radial intensities in the neighborhood of the dark region. We
calculate the intensities by finding the number of rays that cross
in any given area. The potential of the axicon trap is similar to
the potential produced when using a phase plate
\cite{ozeri99,chaloupka99} or a hologram \cite{arlt00}. In all
setups, the height of the potential well forming the trap is
non-uniform. The depth of the trap is determined by the location
with the smallest potential barrier which occurs slightly off axis
in all cases. In the axicon trap, the points of lowest intensity
barrier are close to $z_{\rm far}$ and have an intensity of
$0.20\,I_0$, where $I_0$ is the intensity at lens $L_1$. These
points form a ring of escape avenues for atoms from the trap. The
point of highest intensity is located at $z_{\rm near}$, where
almost all intensity is concentrated and not at $z=D$, where the
intensity is distributed in a ring of diameter $2r_{\rm largest}$
and width $w_{\rm ring}$. These features render the axicon trap
very attractive when combined with gravity, {\it i.e.} when the
laser beam propagates upwards. However, in this work for technical
reasons related to the geometry of the vacuum chamber, the trap is
formed using a light beam that propagates horizontally.

\section{Experimental implementation}

In the laboratory we realized the axicon trap using laser-cooled
metastable xenon atoms. The choice of the atom was arbitrary,
simply that xenon was available for this work. In xenon the
optical dipole transition at $882$\,~nm from the metastable
$6s[3/2]_{2}$ state to the $6p[5/2]_{3}$ state is used for laser
cooling. The lifetime of the lower state is $43$\,s, and is much
longer than the typical time scale of about $2$\,s for each
individual experiment. The apparatus, as well as the slowing and
trapping sequence have been described in detail in Ref.
\cite{xenon}. The infrared laser light for all cooling and
trapping is provided by Ti:sapphire lasers. Briefly, we collect a
few million atoms in the magneto-optical trap which has an rms
radius of $\sigma \simeq 200\,\mu$m. Cooling by optical molasses
further reduces the temperature of the atoms to about $10\,\mu$K.
We record the number of atoms and their arrival times onto a
microchannel plate detector located about $15$\,cm below the
interaction region. In all experiments the time origin is
established by extinction of the light forming the optical
molasses. The atoms reach the detector after about $175$\,ms of
free flight. The spread in arrival times reflects the spread in
velocities of the atoms and can be related to their temperature.

\begin{figure}
\leavevmode \centering \epsfxsize=14cm \epsffile{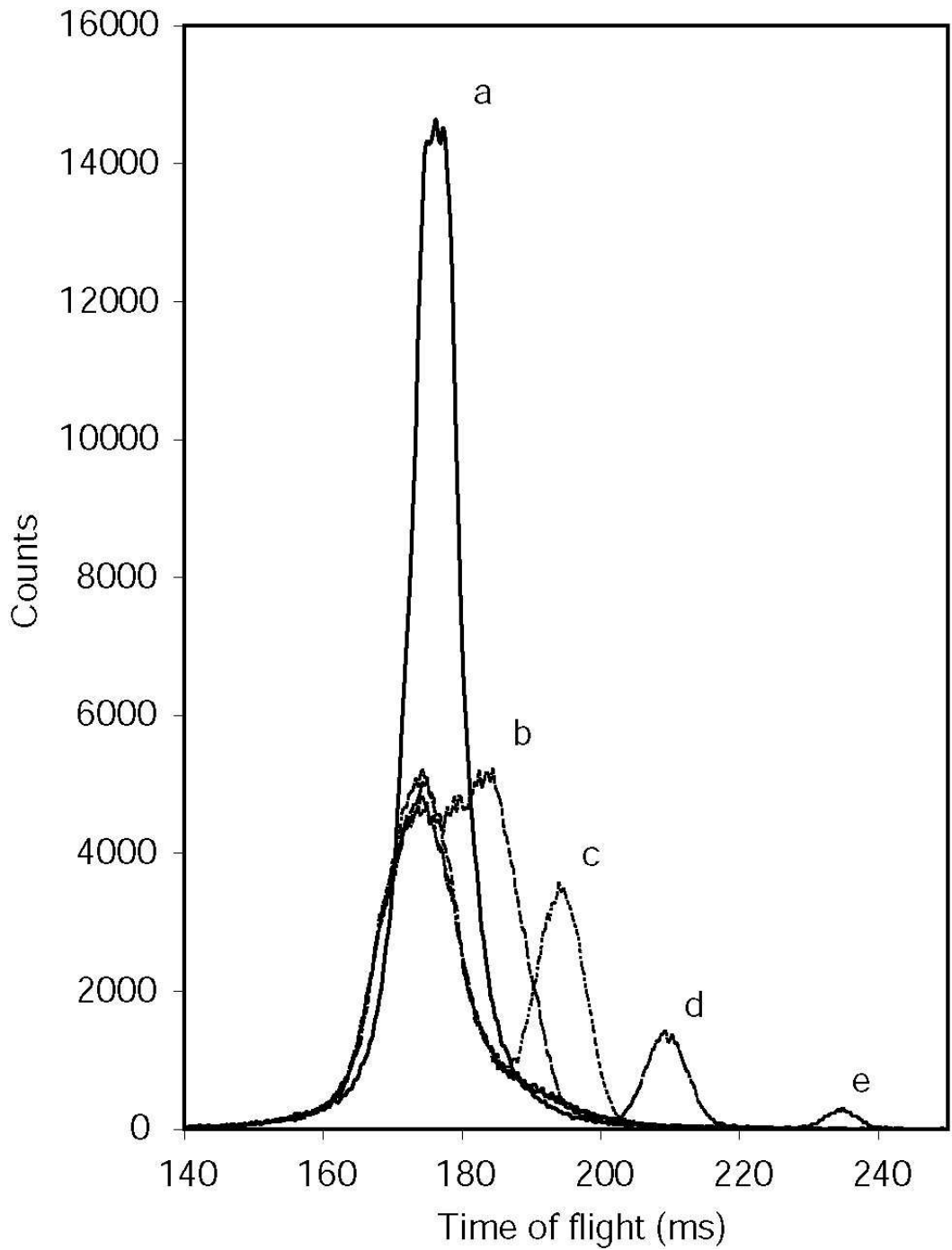}
\caption{Time of flight signals. The curve $a$ is the reference
peak, corresponding to atoms released from the optical molasses.
The other curves are recorded after transferring the atoms into
the axicon trap and holding them for $b:10$, $c:20$, $d:35$, and
$e:60$\,ms respectively. Experimental parameters: laser power
$70$\,mW and detuning $11.7$\,GHz. The curves shown represent
averages over 20 experimental cycles. } \label{TOF}
\end{figure}

We load cold atoms from the magneto-optical trap into the
co-located axicon trap to test it. Light for this trap comes from
a second Ti:sapphire laser and is detuned above resonance by about
$\simeq 2\times 10^{3}\,\Gamma$ from the cooling transition, where
$\Gamma/2\pi = 5.2$\,MHz is the natural linewidth of the upper
level. The optical power in the beam is typically between
$25-75$\,mW and the size of the beam at lens $L_{1}$ is
approximately $8$\,mm in diameter. In the time sequence of the
experiment we switch off the near resonant laser light of the
optical molasses and simultaneously turn on the far detuned light
of the axicon trap. The results for different durations during
which the axicon beam is turned on are shown in Fig.~\,\ref{TOF}.
The first time of flight signal shown is a reference signal (a):
all atoms are released from the optical molasses without being
loaded into the axicon trap. All other signals correspond to
situations in which the axicon trap was kept on for variable
trapping durations $\tau$. A second peak, displaced in arrival
time by $\tau$ from the reference peak, appears in the signal and
represents atoms held in the axicon trap. Atoms that were lost in
the transfer form a peak centered around $175$\,ms, the ballistic
time of flight.

To ensure an efficient loading of the
axicon trap, care is taken to spatially overlap the dark region
in the laser beam forming the axicon trap with the magneto-optical trap.
We achieved transfers of up to $\approx 50$\% of the atoms from the
magneto-optical trap into the axicon trap. It
is difficult to precisely determine the fraction of atoms captured
by the axicon trap. When $\tau \leq 10 $\,ms the
two peaks in the time of flight signal partly overlap, whereas at
larger values of $\tau$ loss of atoms from the axicon trap starts to
become important.

\begin{figure}
\leavevmode
\centering
\epsfxsize=14cm
\epsffile{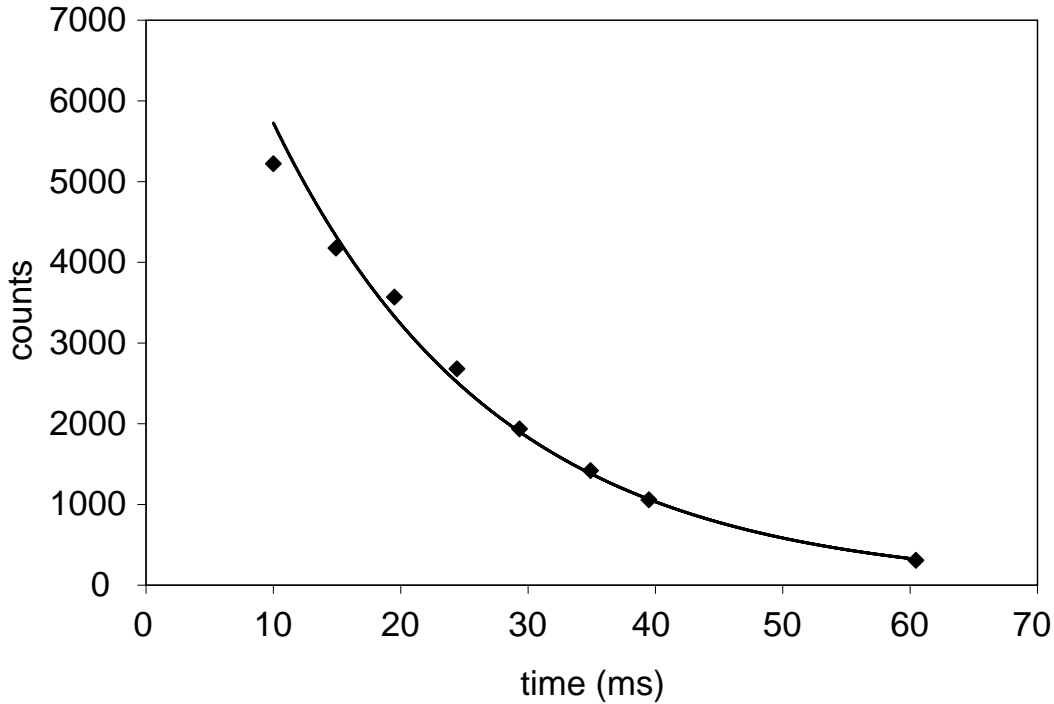}
\caption{Loss of atoms from the axicon trap as a function of
trapping time. Laser power $75$\,mW, detuning $15$\,GHz.
\label{fig8new}}
\end{figure}

We vary the duration of time $\tau$ during which atoms are
held in the axicon trap to determine the lifetime of the
trap. The result of a series of such  measurements is shown in
Fig.~\ref{fig8new}. After about $30$\,ms half of the initially trapped
atoms are still in the trap. This characteristic time strongly
depends on the detuning $\delta$  and the total intensity $I$ of
the laser forming the
trap.  In order to study how the lifetime of the trap is affected by
the detuning we performed a measurement in which we
vary the spontaneous scattering rate ($\propto I/\delta^{2}$), but
keep the height of the potential barrier experienced by the
atoms($\propto I/\delta$) constant. The lifetime of the trap increases
with the detuning, consistent with heating due to spontaneous emission.

We have estimated the number of spontaneous emission events as an
atom with an energy corresponding to 10 $\mu$K climbs the
potential barrier in various locations inside the trap. The recoil
energy for a Xe* atom per scatter is 0.186 $\mu$K. For atoms at
10$\mu$K, the effect of gravity on the energy of the atoms cannot
be neglected. The ramp which gravity superimposes onto the light
potential has a slope of $3.2$\,mK/cm. An atom released on axis at
z=D falls $0.74$\,mm hitting the focal ring in $12.3$\,ms with a
maximum horizontal excursion of $0.3$\,mm while gaining
$230\,\mu$K in kinetic energy. During the fall the atom remains
primarily in the dark region of the trap and scatters about 160
photons per second (70 mW power, 11.7 GHz detuning). The atom,
which now has a kinetic energy of somewhat below $240\,\mu$K, is
repelled by a potential wall of nearly $325\,\mu$K. Atoms released
within an interval of 12 mm, roughly centered at $z=D$ are
repelled by a potential wall of at least $240\,\mu$K. Outside this
range, the atoms can gain enough kinetic energy as they fall down
to jump the potential wall. This confining region is $24\%$ of the
total trapping volume. This emphasizes the importance of a careful overlap
between the region in space where the magneto-optical trap is
formed and the location of the bright ring of the axicon trap.

At this point we can say that our results are consistent with
losses due to gravity and the effect of heating due to spontaneous
scattering \cite{metcalf99} and we can not exclude losses due to
light assisted collisions \cite{collsuppresion}. The effective
volume of the trapping region may be increased by overlapping the
confining regions of two such axicon traps (orthogonal optical
axes) or by directing the optical axis of a single trap
vertically.

\section{Conclusions}
We have demonstrated a new optical arrangement to generate regions
of darkness that we have used to trap laser cooled xenon atoms. We
created this dark volume by placing an axicon lens in the path of
a single laser beam tuned below the atomic resonance. Geometric
optics allows quantitative understanding of the shape and size of
the trapping potential. The dark volume can be as large as
$80$\,mm$^{3}$ and we
 have trapped up to one million atoms in it.

Due to its simplicity the axicon trap may be an attractive tool
for precision experiments. We would like to use this blue-detuned
optical trap for francium atoms on which we perform parity
nonconservation (PNC) measurements. We can extract the PNC
information from precision measurements of electromagnetically
forbidden transitions 
that are investigated in an
environment of well defined handedness.
A far detuned blue optical dipole trap has the advantage
that the externally applied DC
electric and magnetic  fields, as well as the
${\vec k}$ of the laser excitation field are decoupled
from the light field that provides the trap.
We have demonstrated trapping of
$1000$ $^{210}$Fr atoms in a magneto-optical trap \cite{simsarian96},
a number that we have increased by a factor of ten \cite{sprouse}. We are
currently working on improving the number of trapped atoms by
using a double MOT system and thus having a much longer trap
life-time. Radioactive $^{210}$Fr has a lifetime of
approximately
$3$\,min which, when contained in a trap,
is long enough to perform the sequence of
measurements necessary for determining the parity violation signal
\cite{wiemanCs}. With respect to other precision spectroscopy to
test discrete symmetries, Romalis and Fortson \cite{fortson} have
shown that to perform a more sensitive test of time invariance
with trapped atoms in a dipole trap than presently done, it is
necessary to have about $10^{8}$ trapped atoms interrogated for
about $10$\,s.

Francium is the heaviest alkali atom, and thus the fine structure
of the $D$ line is significantly modified by relativistic effects.
The fine splitting between the $D_1$ (817 nm) and the $D_2$ (718
nm) lines of Fr is 100 nm. For all PNC measurements the atoms have
to be prepared in one magnetic substate and this is generally done
with optical pumping. In the sequence of PNC determination, the
m-state will be flipped from $m_{F}=13/2$ to $m_{F}=-13/2$.
Therefore it is important that the light shifts in the dipole trap
be the same for all Zeeman sublevels. In order to minimize the
sensitivity to polarization effects due to the trapping light, the
detuning has to be larger than the fine structure splitting of the
$D$ line \cite{salomon}. This requires $32$\,W of laser power at
$532$\,nm in order to produce an axicon trap a fifth of the size
of the one discussed in this work, i.e. a trap with a ring
diameter of $300\,\mu$m and ring width of $30\,\mu$m. In such a
trap francium atoms cooled to the Doppler limit ($\approx
180\,\mu$K) can be trapped in the dark. In the above
considerations linearly polarized light was employed. When using
circularly polarized light, the laser can be tuned between the
$D_1$ and $D_{2}$ line and the required intensity is much lower.
However, in such a trap, which was demonstrated experimentally for
rubidium atoms \cite{choKorea}, atoms are trapped in the region of
highest intensity, and therefore it is not an appropriate trap for
precision measurements.

We are currently also investigating a new configuration of far
detuned trap, formed by two counter-propagating ``axicon beams''
which permit a much tighter and more symmetric trapping potential.
However, the simplicity of the single beam axicon trap may remain
preferable.

We thank T. Killian for assistance with data acquisition.
S. Kulin acknowledges support from the Alexander von Humboldt
Foundation,
S. Christe and B. Peker received REU support from the NSF.  This
work has been supported by NSF and the Guggenheim Foundation.


\begin{thebibliography}{<27>}
\bibitem{naples} While preparing this manuscript, we became aware of
similar work pursued independently in the group of Tino G at the
University of Naples, Italy. The work there is focussed on making
an axicon trap as small as possible in order to trap cold atoms
from a Bose Einstein condensate.
\end{thebibliography}
\end{document}